# Coupling Mode of Dual-Core Micro Structured Fibers

DEBBAL Mohammed [1], CHIKH-BLED Mohammed [2]

[1]University of Tlemcen, Algeria,
Department of electrical Engineering, Faculty of Technology, Telecommunication Laboratory.
13000 Tlemcen, Algeria, E-Mail: debbal.mohammed@gmail.com

[2]University of Tlemcen, Algeria,
Department of electrical Engineering, Faculty of Technology, Telecommunication Laboratory.
13000 Tlemcen, Algeria, E-Mail: mek_chikhbled@yahoo.fr

*Abstract* – *The photonic crystal fibers (PCF) or air-silica microstructured fibers consist of a periodic array of dielectric transverse. By introducing a defect in this structure, it is possible to guide the light by a photonic bandgap effect, whose properties are different fundamentally from the guide by total internal reflection that takes place in conventional fibers.*
*PCF with two cores have significant potential, and this is one of the main motivations witches led us to approach this theme in this article.*
*Analysis of the inter-core coupling was also necessary to study the problem of crosstalk. Their knowledge is important because it is a preliminary work to the study and understanding of multi-core PCF or an array of guides in the microstructured cladding. It then presents the main results on the effects of beating between the various modes under linear conditions.*

## I. INTRODUCTION

The coupling in PCF can be modeled by two identical waveguides that are brought close to each other so that they can exchange energy with each other through the evanescent part of the fundamental modes of each guide. These evanescent parts depend on the wavelength, which will result in different energy exchanges depending on the wavelength considered.

We consider a PCF consists of two cores (Fig.1). In this case the light is guided by total reflection (Fig.2). It periodically passes from one core to another (Fig.3).

This PCF is achieved by introducing two defects (two cores) in the microstructured cladding.

Figure 1 shows a cross section of this PCF, characterized by a diameter of the core "Dc", a hole diameter "d", and a hole spacing "Λ". Figure 2 shows the principle of the guide in such fiber.

A part of the signal at the input of the core (a) appeared in the output of the core (b). With the transfer of power between two cores of PCF is periodically (Fig.3).

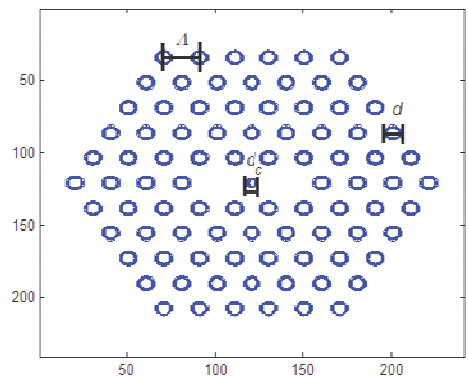

Fig. 1. Cross section of dual core PCF

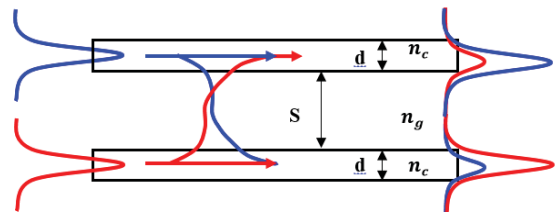

Fig. 2. Mode coupling between two guides

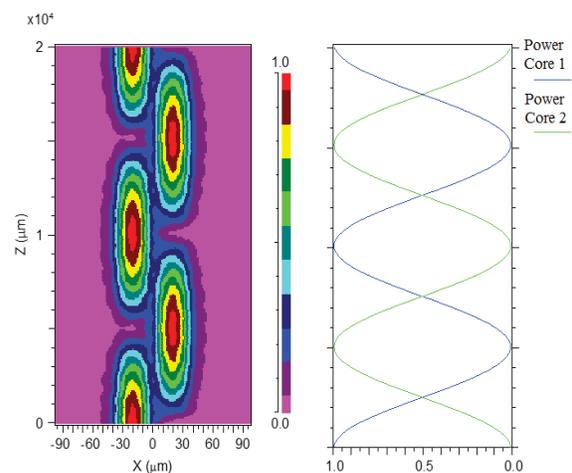

Fig. 3. Power transfer between cores.

## II. DETERMINATION OF THE COUPLING COEFFICIENTS AND THE COUPLING LENGTH

This part is dedicated to the calculation of coupling lengths and coefficients for PCF with dual-cores according to the geometric parameters [1 - 7].

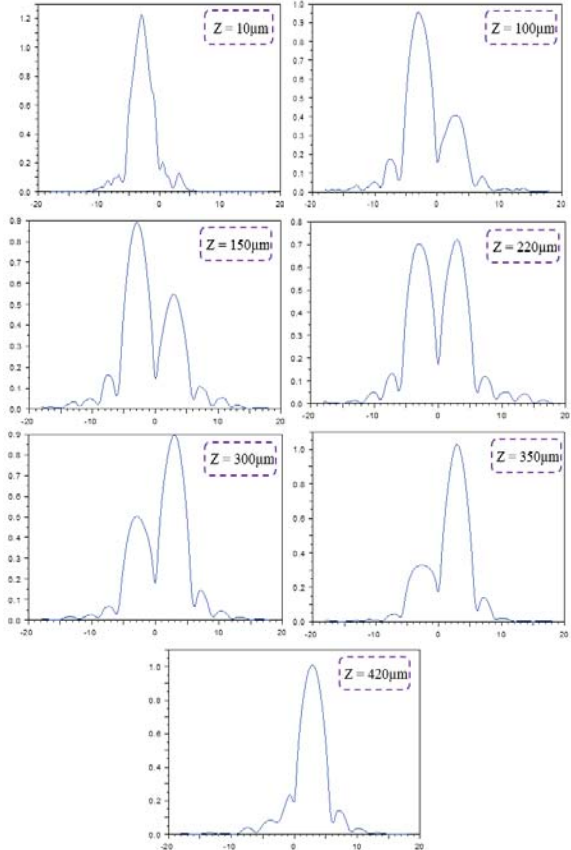

Fig. 4. Field amplitude in dual-cores PCF for different distances Z.

PCF consists of dual-core with a diameter dc = 0.6μm. The space between air holes of the microstructured cladding (center-to-center spacing) Λ = 3µm.

Figure 4 shows the amplitude of the electric field E in the dual core PCF for different distances Z. We can observe the phenomenon of inter-cores beat, which is to say that there is an exchange of power between the first core and the second core. At the beginning of the propagation, the field in the first core is maximal, approximately 1.3, which will decrease during the propagation. Reverse against in the second core, the field increases to a distance of 220μm to equalize with the first core field.

The distance for which the amplitude of the second core is maximum is defined as the coupling length Lc.

Theoretically this distance is obtained from the propagation constants of the even and odd modes resulting from the coupling between the evanescent fields from each core.

The analytical expression for this parameter is given by [8]:

$$L_c = \frac{\pi}{\beta_{even} - \beta_{odd}} = \frac{\lambda}{2(n_{even} - n_{odd})}$$

βeven, βodd are the propagation constants of the even and odd modes of propagated mode.

The coupling coefficient is deducted from the coupling length Lc:

$$C_{x/y} = \frac{\pi}{2L_{c_{x/y}}}$$

## III. COUPLING COEFFICIENTS AND THE COUPLING LENGTH DEPENDING OF GEOMETRIC PARAMETERS

In order to study the influences of the geometric parameters of the dual core PCF on the length coupling and the coefficient coupling we are going to proceed to a variation of the various geometric parameters.

Figure 5 shows the variations of the coupling length versus the hole diameter of the microstructured cladding, for a core diameter dc = 0.6μm, various spacing Λ equal to 2, 3, 3.5 and 4μm with a wavelength λ = 1.55μm.

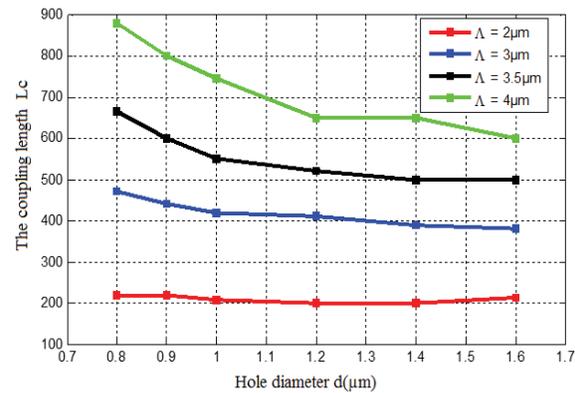

Fig. 5. Hole diameter dependence of coupling length for a PCF coupler, taking the operating pitch (Λ) as a parameter.

Figure 5 shows that the coupling length is directly dependent on the geometrical parameters of PCF. It is maximal, and the value achieved is 880μm for a spacing Λ = 4μm and d = 0.8μm. By reducing the spacing Λ between holes and the hole diameter, the coupling length decreases. It is approximately 213μm for Λ = 2μm and d = 1.6μm.

The structures of the large hole diameter are more susceptible to decoupling because the mode is confined to one of the cores. These structures make it possible to obtain coupling lengths significantly lower those achieved using conventional couplers. This coupling length can be adjusted by varying the geometric parameters of the structure.

Figure 6 shows the variations of the coupling coefficient as a function of hole diameter for a diameter core dc = 0.6μm.

The distance between the holes Λ varying from 2, 3, 3.5 to 4µm with a wavelength λ = 1.55µm. The coupling coefficient is inversely proportional to the hole diameter. It is lower for a maximal hole diameter, and a large spacing between holes.

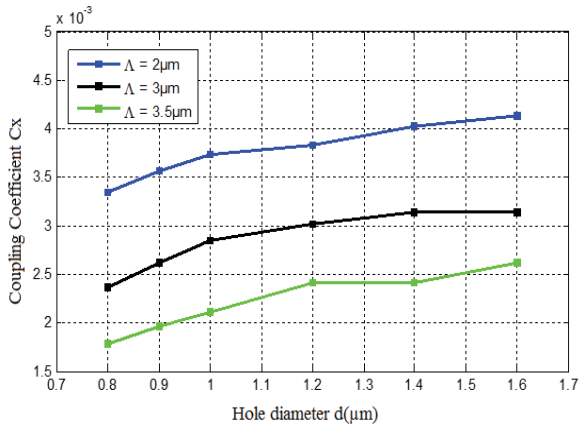

Fig. 6. Hole diameter dependence of coupling coefficient for a PCF coupler, taking the operating pitch (Λ) as a parameter

## IV. COUPLING COEFFICIENTS AND THE COUPLING LENGTH DEPENDING OF THE CENTRAL HOLE DIAMETER

Now we will study the influence of the central hole diameter Dc which separates the two cores on the length and the coefficient coupling.

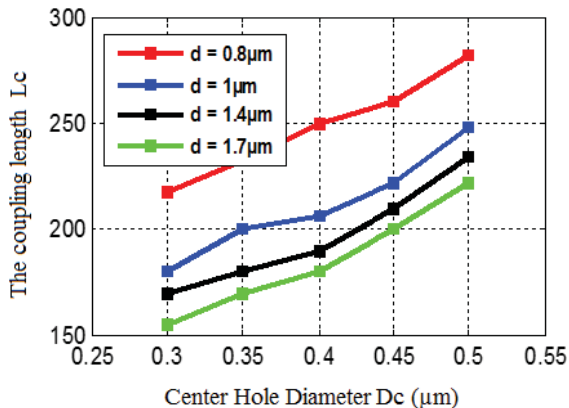

Fig. 7. Central hole diameter dependence of coupling length for a PCF coupler, taking the operating diameter (d) as a parameter

The results are shown in Figure 7 and 8 which translates the variations of coupling length and coupling coefficient depending on the central hole diameter of a hole spacing Λ = 2.5µm. The hole diameter varies d = 0.8, 1, 1.4 and 1.7µm. The simulations were performed at a wavelength λ = 1.55µm.

Note in Figure 7 that the central hole diameter plays an important role in the coupling length, which becomes maximal when increasing the diameter dc. It reaches a value of approximately 155µm for a diameter dc = 0.3µm and d = 1.7µm.

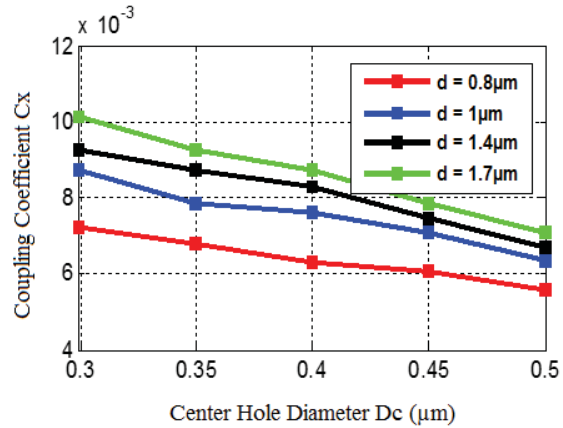

Fig. 8. Central hole diameter dependence of coupling coefficient

The difficulty lies in making a dual core PCF for the small dimensions of the central hole. R.Buczynski and al have made a photonic structure with the following parameters: Λ = 1.81µm, d = 0.61µm (d/Λ = 0.34) and dc = 0.45µm [9] (Fig.9). Thus require proper selection central hole diameters practically feasible.

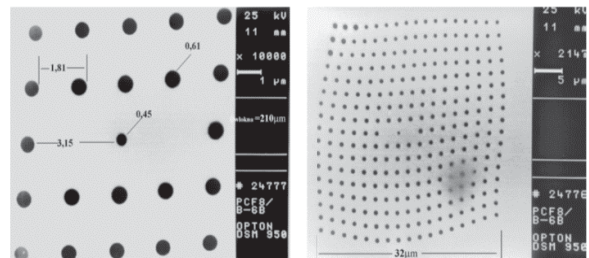

Fig. 9. Dual core PCF, fiber diameter 220µm, Λ = 1.81µm, hole diameter d = 0.61µm (d/Λ = 0.34) and the center hole diameter dc=0.45µm [9].

## V. CONCLUSION

A detailed study on the mode coupling in a dual core PCF.

These studies aimed at first to understanding the mode coupling in terms of transfer of power from one core to another, and which is characterized by a phenomenon of interference or beat of these modes called supermodes.

This brings us to the phenomenon of crosstalk resulting from the inter-core coupling.

My work has focused on the coupling coefficient in dual core PCF. It is a preliminary work to the study of more complex structures such as multi-core PCF.

It then presents the main results on the effects of beating between the various modes in the linear regime.

From the characteristics of mode coupling for dual-core PCF, several applications can be presented as the wavelength demultiplexer and polarization splitter. This is key components for optical telecommunication systems.